# Evidence for Three-component Interlayer Coherent Exciton Condensation

Subi Du[1,2], Xiaohan Zhang[1,2], Hongxi Song[3], Siyu Fan[1,2], Yiduo Wang[1], Zhenyu Wang[1,2], Song Liu[4], Kenji Watanabe[5], Takashi Taniguchi[6], Jiangping Hu[1,2]*, Yang Xu[1,2]*

[1]Beijing National Laboratory for Condensed Matter Physics, Institute of Physics, Chinese Academy of Sciences, Beijing, China.
[2]School of Physical Sciences, University of Chinese Academy of Sciences, Beijing, China
[3]Department of Physics, University of Science and Technology of China, Hefei, China.
[4]Institute of Microelectronics, Chinese Academy of Science, Beijing, China
[5]Research Center for Functional Materials, National Institute for Materials Science, Tsukuba, Japan
[6]International Center for Materials Nanoarchitectonics, National Institute for Materials Science, Tsukuba, Japan
[7]New Cornerstone Science Laboratory, Institute of Physics, Chinese Academy of Sciences, Beijing, China
*Email: jphu@iphy.ac.cn; yang.xu@iphy.ac.cn

## Abstract

**Increasing the number of internal components in a quantum many-body system can host collective orders inaccessible to simpler settings. Quantum Hall bilayers provide a canonical realization of interlayer exciton condensation, yet extending such coherence across three independently addressable electronic fluids has remained elusive. Here we report evidence for three-component interlayer-coherent exciton condensation in triple-layer graphene system. Using Rydberg excitons in an adjacent $WSe_2$ monolayer as a layer-sensitive optical probe, we resolve interaction-induced incompressibility at zeroth-Landau-level crossings for all three pairwise layer combinations, establishing top–middle, middle–bottom and top–bottom exciton condensate channels within the same device. Independent control of displacement field and interlayer bias continuously tunes these pairwise states towards a regime where Landau levels from all three layers approach simultaneous degeneracy. At their convergence, incompressibility persists while the exciton energy and spectral weight evolve smoothly between the pairwise limits, suggesting coherent participation of all three layers in a single three-component state. More broadly, the ability to independently control layer potentials and engineer interlayer interactions establishes multilayer graphene as a programmable synthetic dimension for exploring higher-component quantum Hall order and simulating strongly correlated quantum matter.**

## Introduction

When internal flavors such as spin, valley or layer are present, electron interactions in the quantum Hall (QH) regime can transform a discrete Landau-level (LL) spectrum into a hierarchy of correlated many-body ground states [1-3]. A foundational framework for such multicomponent QH states is provided by the Halperin wavefunctions, which encode correlations both within and between different flavor components [1,2]. The layer degree of freedom is distinctive because it combines an internal quantum number with real-space separation, allowing the relative strength of intra- and interlayer Coulomb interactions to be controlled geometrically [4,5]. In a bilayer at total filling $\nu_{\mathrm{tot}} = 1$, the Halperin (111) state provides the canonical microscopic description of the interlayer-correlated QH state [1,6-11]. In the ideal limit of negligible interlayer tunnelling and vanishing layer separation, the system has an SU(2) layer-pseudospin symmetry. Finite separation $d$ introduces an easy-plane anisotropy, reducing the symmetry to a relative U(1) [6-8]. At sufficiently small $d/l_{\mathrm{B}}$ (where $l_{\mathrm{B}}$ is the magnetic length, setting the characteristic intralayer separation of quasiparticles within a LL), interlayer exchange stabilizes an easy-plane pseudospin ferromagnet, equivalently an exciton condensate (EC) of electrons and holes in separate layers [5-8,12]. The interlayer coherence spontaneously breaks the U(1) symmetry by selecting a single interlayer phase, giving rise to a charge-gapped coherent state identified experimentally through tunnelling, drag, and counterflow transport measurements [9-11,13-19].

Extending this physics to three layers introduces two independent relative phases and, in the ideal limit, an approximate SU(3) symmetry in layer space [20,21]. Finite layer separations introduce capacitive and exchange anisotropies, but when tunnelling is negligible the two relative U(1) symmetries remain. The system can consequently support coherence in each layer pair (i.e., top–middle, middle–bottom and top–bottom) as well as a fully coherent state involving all three layers, in which both relative U(1) symmetries are spontaneously broken and two neutral collective phase modes emerge [20-22]. Theories have further predicted correlated and topological phases that go beyond the paradigm of bilayer physics [23-27], including anyon-exciton superfluidity [23], multicomponent fractional QH states with irrational charge distributions [24], and chiral surface phases in the many-layer limit [25,26]. Experimental access to this higher-component regime, however, has remained limited [28-32]. Earlier GaAs triple quantum wells established high-mobility three-layer electron systems, but buried-layer geometries, finite subband hybridization, and limited layer-selective control and readout made it difficult to resolve the contribution of each layer or to identify an unambiguous three-component coherent state [28-31].

Graphene-based van der Waals (vdW) heterostructures provide a route to overcome these limitations [18,19]. Monolayer graphene hosts a fourfold-degenerate zeroth Landau level (ZLL) whose degeneracies can be fully lifted by interactions [33-40], while atomically thin hBN spacers suppress single-particle tunnelling yet preserve strong interlayer Coulomb coupling [18-19]. Gates and individual electrical contacts further enable precise control of the carrier density and electrostatic potential in each layer. Here we combine this layer-addressable graphene platform with Rydberg exciton spectroscopy in a nearby monolayer $WSe_2$ sensor (see the full device structure in Extended Data Fig. 1) [41-48]. Because the optical response depends on the distance to each graphene layer, this approach further enables layer-sensitive access to the LL spectrum without requiring separate transport readout from every layer. We first demonstrate the method in double-layer graphene (DLG), identifying layer-resolved ZLL crossings and incompressible states consistent with two-component interlayer coherence. We then extend it to triple-layer graphene (TLG) and observe incompressible states associated with all three pairwise layer combinations. We further map a $\nu_{\mathrm{tot}} = 1$ phase diagram in which interlayer bias and displacement field continuously redistributes charge among the three layers, driving continuous evolution between layer-polarized QH states, pairwise interlayer-coherent states and a three-component interlayer-coherent regime near simultaneous LL degeneracy.

**Layer-resolved LL crossings and EC in DLG**

We first use DLG to establish layer-resolved optical detection of QH states and benchmark the signatures of interlayer coherence. The device consists of two monolayer graphene sheets separated by an approximately 2 nm hBN spacer, with a monolayer $WSe_2$ placed above the top graphene layer as an optical sensor (Fig. 1a). Because the $WSe_2$ is closer to the top graphene layer, its long-range Coulomb coupling to the graphene electronic response is stronger for the top layer than for the bottom layer, providing intrinsic layer sensitivity (see more discussions in Methods). The total filling factor $\nu_{\mathrm{tot}}$ and displacement field $D$ are independently controlled by the top and bottom graphite gates. At $B$ = 9 T, the interlayer separation is substantially smaller than the magnetic length ($d/l_B \approx 0.23$), placing the two graphene layers in the strong coupling regime [7,9,10,18,19].

The optical response provides several complementary views of electronic states in DLG. Figure 1b shows the reflection contrast ($\Delta R/R_0$) spectrum (see Methods) across the graphene ZLL at $D$ = 80 mV/nm. Interactions lift the fourfold spin–valley degeneracy of the ZLL in each graphene layer, resolving it into four symmetry-broken sublevels [33-40]. When both layers are at integer fillings, the whole system becomes strongly incompressible, suppressing its low-energy charge response and thereby reducing the screening of the $WSe_2$ excitons. The resulting change in the excitonic response therefore provides an optical signature of QH incompressibility [44,45,48]. Under these conditions, a well-developed Rydberg series can be resolved from the 2s state to at least the 10s state. By contrast, when either graphene layer is partially filled, its enhanced charge response strongly modifies the excitonic spectrum, suppressing the oscillator strength of the neutral Rydberg-exciton resonances and redistributing their spectral weight. Coupling to the electronic excitations of the partially filled graphene Landau levels further gives rise to additional interlayer exciton-polaron features [45], which are more readily resolved for the lower Rydberg states under the contrast conditions of the present measurements.

At $D$ = 80 mV/nm, the ZLLs of the two graphene layers are energetically offset and fill sequentially, with the top-layer filling factor $\nu_{\mathrm{t}}$ first swept from −2 to 2, followed by the bottom-layer filling factor $\nu_{\mathrm{b}}$ from −2 to 2. This layer dependence is particularly evident in fixed-energy color maps in the $\nu_{\mathrm{tot}}$–$D$ plane (Fig. 1c). Maps taken through the 2s ($E$ = 1.813 eV) and 3s ($E$ = 1.826 eV) resonances resolve distinct features associated with the two graphene layers, including their symmetry-broken ZLL sublevels, incompressible QH gaps, and interlayer LL crossings. The $\Delta R/R_0$ amplitudes at these energies exhibit a pronounced layer dependence. Features associated with the top graphene layer generally exhibit stronger contrast because of its closer proximity to the $WSe_2$ sensor, whereas those from the bottom layer carry weaker spectral weight modifications [43,46,48]. For example, at $E$ = 1.813 eV, the stronger set of LL trajectories (red regions) is assigned to the top graphene layer, whereas the weaker set (white regions) originates from the bottom layer. Varying $D$ shifts the four ZLL sublevels of the two layers relative to one another, generating the full 4 × 4 network of 16 interlayer LL crossings within the ZLL manifold. Resolving this full crossing network demonstrates the layer selectivity and spectral resolution of the exciton-sensing approach. With increasing principal quantum number ($N$), however, both the exciton-polaron response and the layer contrast progressively weaken. At the 8s resonance ($E$ = 1.864 eV), the two layers acquire similar optical weights and are no longer readily distinguished. Thus, different Rydberg exciton states provide complementary sensitivity to the total and layer-resolved electronic response of the DLG. The energy-dependent layer selectivity is further illustrated over a broader spectral range in Extended Data Fig. 2.

The resulting ZLL sublevel crossing pattern is captured by an electrostatic model that incorporates the interaction-induced QH gaps and negative compressibility of the individual graphene ZLLs (see Methods; Fig. 1d). Away from the interlayer crossings, for example at $D$ = 80 mV/nm, the $D$ field separates the chemical potentials $\mu_{\mathrm{t}}$ and $\mu_{\mathrm{b}}$ of the two layers, leading to sequential evolution of $\nu_{\mathrm{t}}$ and $\nu_{\mathrm{b}}$ (left panels of Fig. 1e). At $D$ = 0, by contrast, corresponding symmetry-broken ZLL sublevels in the two layers become degenerate ($\mu_{\mathrm{t}} = \mu_{\mathrm{b}}$, right panels of Fig. 1e). Charge is then distributed equally between the two layers, with $\nu_{\mathrm{t}} = \nu_{\mathrm{b}} = \nu_{\mathrm{tot}}/2$ at the four crossings. The calculation reproduces the full network of 16 top–bottom ZLL crossings, but does not include interaction-driven gap opening between the layers.

The interlayer interaction-induced incompressibility is clearly seen along the $D$=0 trajectory (Fig. 1h). In addition to the regular QH gaps at even $\nu_{\mathrm{tot}}$, where both layers are integer filled, sharp additional features emerge near the 2s exciton-polaron energy ($E$ = ~1.810 eV) precisely at $\nu_{\mathrm{tot}}$=−3, −1, 1 and 3 (highlighted by orange arrows in Fig. 1h). These four states occur at the ZLL crossings where the corresponding ZLL sublevels in the two layers are simultaneously half filled. For example, $(\nu_t, \nu_b)$ = (1/2, 1/2) at $\nu_{tot}$ = 1 and (3/2, 3/2) at $\nu_{tot}$= 3. For $d << l_B$ and negligible tunnelling, interlayer exchange favors a coherent superposition of the two degenerate layer states, producing an easy-plane pseudospin ferromagnet, equivalently an interlayer EC (Fig. 1f,g) [5–10,18,19]. The additional incompressibility observed at these crossings is therefore consistent with an interaction-induced excitonic gap.

More generally, comparable incompressible features are observed not only at these four balanced crossings but also at the remaining 12 crossings between different symmetry-broken ZLL sublevels (see the color map taken at $E$ = ~1.810 eV in Fig. 1c). At the centers of these crossings, the two layers carry unequal filling factors that differ by an integer number [e.g., $(\nu_t, \nu_b)$ = (1/2, 3/2)], although both active crossing sublevels remain half filled. Additional examples across both even and odd $\nu_{tot}$ for $D > 0$ are shown in Extended Data Fig. 3. A particularly distinct case occurs at $\nu_{tot}$ = 0 [e.g., $(\nu_t, \nu_b)$ = (1/2, −1/2)], where the electron-like and hole-like contributions of the two layers carry opposite sign and the net charge Chern number vanishes. The interaction-induced gap in this case realizes a zero-Chern exciton insulator rather than a finite-Chern QH EC. Furthermore, detuning $D$ away from centers of these 16 crossings at fixed integer $\nu_{tot}$ continuously transfers charge between the two layers while the interaction-induced incompressibility persists over a finite range, demonstrating that the coherent states can accommodate substantial charge imbalance from half-filled ZLL sublevels.

**Pairwise interlayer coherence in TLG**

We next extend the optical detection scheme to a TLG device, where the layer degree of freedom expands from two to three components. The device contains top, middle and bottom monolayer graphene (denoted as t-Gr, m-Gr, and b-Gr, respectively) sheets separated by thin hBN, with the $WSe_2$ sensor positioned above m-Gr (Fig. 2a). For the measurements in Fig.2, a small bias voltage $V_{bias}$ is applied to t-Gr, while m-Gr and b-Gr are grounded. This bias compensates the residual electrostatic imbalance and restores a more symmetric ZLL crossing pattern. The different sensor–graphene distances produce different optical weights for the three layers. This is particularly clear in the phase diagram obtained near the 2s exciton resonance (Fig. 2b), where three families of ZLL trajectories can be distinguished and assigned to the m-Gr (dark red), t-Gr (light red), and b-Gr (white), respectively. Their progressively weaker optical contrast reflects the increasing effective distance from the $WSe_2$ sensor. By comparison, the more spatially extended 3s exciton exhibits weaker layer selectivity, such that the top- and bottom-layer features acquire more similar optical weights and are less readily distinguished (Fig. 2c; see also Extended Data Fig.4). The state-dependent spatial sensitivity of the Rydberg excitons therefore enables the electronic responses of three closely spaced graphene layers to be disentangled using a single optical sensor.

The experimentally resolved crossing network agrees well with the calculated electrostatic evolution of the three layer-resolved ZLLs (Fig. 2d). In contrast to the bilayer, the triple-layer system supports three distinct pairwise crossing channels: top–middle, middle–bottom and top–bottom, yielding in total $4 \times 4 \times 3 = 48$ crossings. When ZLL sublevels from two layers become degenerate while the third layer remains at an integer filling, the active pair forms an effective two-component QH subsystem embedded within the triple layers. Interlayer exchange can then stabilize a pairwise EC, while the third layer remains an inert incompressible background [7,20,21]. The three possible condensates therefore correspond to top–middle, middle–bottom and top–bottom coherence, respectively (Fig. 2e) [20,21].

Signatures consistent with all three pairwise condensates are directly visible in the optical spectra. Along the representative displacement field cut at $D$ = 20 mV/nm (Fig. 2f), which intersects five pairwise ZLL crossings, additional red-shifted incompressible features appear whenever two of the three layers are at half-integer filling while the third remains at integer filling, for example at $(\nu_{\mathrm{t}}, \nu_{\mathrm{m}}, \nu_{\mathrm{b}}) = (2, 1/2, 3/2)$. Their layer-resolved configurations identify top–middle (marked by the dark blue arrow), middle–bottom (light blue arrow) and top–bottom (white arrows) coherent states, closely resembling the excitonic gaps established in the DLG. Notably, the spectral shifts and optical strengths differ slightly among the three pairwise pairing channels. These differences are consistent with variations in the effective interlayer Coulomb interaction arising from the unequal geometric separations and screening environments, together with the layer-dependent coupling to the $WSe_2$ sensor [7,20,21,46,48]. Such differences can lead to different interaction-induced gap scales, although the optical red shift itself is not a direct measure of the EC gap. In particular, the top–bottom state, for example $(\nu_{\mathrm{t}}, \nu_{\mathrm{m}}, \nu_{\mathrm{b}}) = (1/2, 0, 1/2)$, demonstrates that pairwise EC can persist even between the two most widely separated graphene sheets. The observation of interaction-induced incompressibility in all three pairing channels therefore establishes the full set of pairwise EC states within a single TLG device and provides the two-component reference states needed to identify the higher-component regime below. Additional examples of three pairwise pairing channels are shown in Extended Data Fig. 5.

**Signatures of three-component interlayer QH coherence**

Having established the three pairwise channels separately, we next use the additional electrostatic degree of freedom provided by the interlayer bias $V_{\mathrm{bias}}$ to control their relative alignment and the associated interlayer charge transfer. We focus on $\nu_{\mathrm{tot}}$ = 1, where varying $D$ redistributes a fixed total LL filling among the three graphene sheets, while $V_{\mathrm{bias}}$ applied directly to t-Gr shifts its relative electrochemical potential. The resulting $D$-$V_{\mathrm{bias}}$ phase diagram exhibits a characteristic tree-like branching structure (Fig. 3a). Each blue domain corresponds to a regular incompressible QH configuration in which all three layers are integer filled, such as $(\nu_t, \nu_m, \nu_b) = (2, 1, -2)$ and $(2, 0, -1)$. The red branches between adjacent integer-filled domains mark pairwise LL crossings with different contrasts, where charge is transferred between two layers while the third remains incompressible. These branches are associated with the top–middle, middle–bottom and top–bottom crossing channels. Most importantly, the three channels can converge at a central junction, where the relevant ZLLs from all three layers simultaneously approach degeneracy, defining a three-layer LL crossing. As $V_{bias}$ is varied, the three pairwise crossing branches continuously move relative to one another and can be driven into or away from this common junction. The corresponding electrostatic calculation reproduces both the branching topology and its evolution with bias (Fig. 3b), demonstrating that $V_{bias}$ provides a controlled route from isolated pairwise crossings to the three-component degeneracy regime (denoted by yellow triangles).

We examine this evolution in detail by sweeping $\nu_{tot}$ from 0 to 2 and the $D$ over ranges that traverse the three LL-crossing branches connecting the integer-filled configurations $(\nu_t, \nu_m, \nu_b) = (1, 0, 0)$, $(0, 1, 0)$, and $(0, 0, 1)$ at three fixed values of $V_{bias}$ (from top to bottom row, $V_{bias}$ =−12, −2.6, and 6 mV, respectively; Fig. 3c–e). We visualize this evolution with complementary $\Delta R/R_0$ maps at 1.809 eV, near the 2s resonance (Fig. 3c), and 1.804 eV, near the 2s exciton-polaron resonance (Fig. 3d), which emphasize different layer contributions, together with the corresponding calculated crossing maps (Fig. 3e). At $V_{bias} = -12$ mV, the crossing structure near the $\nu_{tot} = 1$ crossing is dominated by the top–bottom (tb) channel (highlighted as the diamond domain enclosed by dashed curves in the upper panel of Fig. 3e), while the middle layer remains incompressible at $\nu_m = 0$. Increasing $V_{bias}$ to −2.6 mV shifts the middle-layer ZLL into the same region, bringing the active ZLL sublevels of all three layers close to simultaneous degeneracy. The three pairwise crossing branches consequently converge into a common three-layer crossing region (highlighted by the yellow triangle and denoted as tmb in the middle panel of Fig. 3e). Upon further increasing $V_{bias}$ to 6 mV, the middle-layer ZLL moves through the original top–bottom (tb) crossing, causing the three-layer crossing to separate again into distinct top–middle (tm) and middle–bottom (mb) pairwise branches on opposite sides in $D$. The evolution from a top–bottom crossing, through a three-layer crossing, and finally into two separated pairwise crossings is consistently observed at both probe energies and reproduced by the electrostatic calculation. The continuous evolution of the simulated zoomed-in ZLL crossing map is further shown in Supplementary Video 1, directly visualizing how the pairwise crossing regions move, merge into a common three-layer crossing region, and separate again as $V_{bias}$ is varied. This controlled convergence provides a well-defined structure in which to test whether the simultaneous participation of all three layers supports a three-component coherent state continuously connecting the pairwise condensate limits.

We finally compare the correlated states encountered with sweeping $\nu_{tot}$ trajectories at different $V_{bias}$ and $D$. Figure 4a–d shows spectra through the four representative trajectories indicated by the white dashed lines in Fig. 3c. At $V_{\text{bias}} = -12$ mV and $D = -1.1$ mV/nm, the $\nu_{\text{tot}} = 1$ incompressible feature corresponds to a top–bottom pairwise state with layer fillings (1/2,0,1/2). At $V_{\text{bias}} = 6$ mV, the same total filling instead supports separately resolved middle–bottom (at $D = 0.4$ mV/nm) and top–middle (at $D = 23.4$ mV/nm) coherent states, corresponding to (0,1/2,1/2) and (1/2,1/2,0), respectively. These three configurations therefore provide experimentally calibrated two-component reference states within the same device. At the intermediate bias $V_{\text{bias}} = -2.6$ mV and $D = 5.6$ mV/nm, an incompressible feature emerges at $\nu_{\text{tot}} = 1$ precisely where the relevant ZLL sublevels from all three layers simultaneously approach degeneracy (orange star in Fig. 4b). Rather than appearing as an isolated additional state, this feature connects continuously to the neighboring pairwise coherent states as the layer populations are redistributed by $V_{\text{bias}}$ and $D$ (See Fig. 3a and 3b).

This continuous evolution is also reflected in the $WSe_2$ excitonic response. Figure 4e compares the spectra taken at $\nu_{\text{tot}} = 1$ for the top–bottom, middle–bottom, top–middle and three-layer configurations. Both the 2s resonance energy and its spectral weight vary systematically with the participating graphene layers, reflecting the layer-dependent charge response sensed by the exciton. The three pairwise states define characteristic optical responses associated with the three two-component limits. At the three-layer crossing, both the 2s resonance energy and spectral weight take intermediate values relative to these pairwise reference states (Fig. 4f, Extended Data Fig.7), revealing a smooth evolution of the optical response as the system is tuned through the three-layer degeneracy. This interpolation is consistent with a continuous redistribution of layer participation within the coherent state among all three layers, rather than a transition between spectroscopically isolated pairwise states. Together with the interaction-induced incompressibility at the simultaneous three-layer crossing, these observations provide evidence for a three-component interlayer-coherent EC state connecting the three pairwise EC states.

**Conclusion**

In summary, we have established layer-resolved optical spectroscopy as a probe of multicomponent QH states in graphene heterostructures. In DLG we resolve the full network of interlayer ZLL crossings and identify interaction-induced incompressible states associated with two-component interlayer ECs. Extending this approach to TLG reveals all three pairwise coherent channels and their continuous evolution toward near-simultaneous ZLL degeneracy. At their convergence, incompressibility persists and the optical response evolves smoothly between the pairwise limits, consistent with a three-component interlayer-coherent QH EC. By extending our device architecture to many independently biased layers, the layer degree of freedom can form a programmable synthetic dimension, with electrically tunable onsite energies and geometrically engineered Coulomb and tunnelling couplings, opening a route to quantum simulation of higher-component and strongly interacting lattice models.

**Acknowledgements**

We thank J. Ye and Z. Zhu for valuable discussions. This work was supported by the National Natural Science Foundation of China (Grant Nos. 92476108 and 12174439) and the National Key R&D Program of China (Grant No. 2021YFA1401300). J. Hu was supported by New Cornerstone Science Foundation. The growth of hBN crystals was supported by the Elemental Strategy Initiative of MEXT, Japan, and CREST (JPMJCR15F3), JST.

**Author Contributions**

Y.X. and S.D. conceived the experiment and co-wrote the manuscript. S.D. fabricated the devices and performed the measurements and data analysis. X.Z. and H.S. performed the simulations. K.W. and T.T. provided the hBN crystals. L.S. provided the $WSe_2$ crystals. J. Hu provided the theoretical supports. All authors discussed the results and commented on the manuscript.

## Methods

### Device fabrication

Monolayer graphene, monolayer $WSe_2$, hBN, and few-layer graphite were mechanically exfoliated from bulk crystals onto $SiO_2$(285 nm)/Si substrates. Flakes with appropriate thickness and lateral size were identified by optical contrast. The heterostructures were sequentially picked up and stacked using a polymer-based dry transfer process [52,53]. A monolayer $WSe_2$ was incorporated into the stack as an optical sensing layer and also served as part of the dielectric stack. Few-layer graphite flakes were used as top and bottom gate electrodes. In the DLG and TLG devices, adjacent monolayer graphene sheets were separated by thin hBN spacers whose thicknesses were determined by atomic force microscopy. After assembly, the completed heterostructures were released onto $SiO_2$/Si substrates with pre-patterned Cr/Au electrodes providing electrical contacts to the individual graphene layers and graphite gates. Optical images of the constituent flakes and the final TLG device, and the complete device structure are shown in Extended Data Fig. 1. Voltages applied to the top and bottom graphite gates using Keithley 2400 source-measure units were used to control the carrier density ($n_{\mathrm{tot}}$) and displacement field ($D$), while the individual graphene layers could be independently grounded or voltage-biased as required.

### Optical measurements

Optical measurements were performed with the devices mounted in a closed-cycle optical cryostat (attoDRY2100) with a base temperature of 1.8 K and a perpendicular magnetic field up to 9 T. A broadband white light source was coupled into a single-mode fiber and collimated with a ×10 objective. For magneto-optical measurements, circularly polarized light was generated using a Glan–Taylor polarizer and an achromatic quarter-wave plate. The beam was focused onto the sample by a low-temperature-compatible apochromatic objective with a numerical aperture of 0.82, resulting in a spot size of approximately 1 μm on the device. To minimize additional optical heating, the illumination was spectrally filtered using a 650-nm long pass filter and a 700-nm short pass filter, restricting the incident light to the spectral range containing the $WSe_2$ Rydberg excitons of interest. The incident optical power on the sample was maintained at the level less than tens of nanowatts. The reflected light from the sample was collected by the same objective and directed into a spectrometer for analysis. The optical response was presented in terms of reflection contrast, defined as $\Delta R/R_0 = (R - R_0)/R_0$, where $R$ and $R_0$ denoted the reflected spectra from the sample and from a nearby region without monolayer $WSe_2$, respectively.

**Generalization from two- to three-component QH coherence**

The interlayer-coherent states discussed in this work can be understood as a multicomponent generalization of the conventional bilayer QH exciton condensate [7,9-11,20,21]. Consider a set of nearly degenerate layer-resolved LL sublevels with a total occupation of one electron per magnetic flux quantum after fully filled or empty sublevels are treated as inert backgrounds. In a bilayer, the active electronic state at each guiding center can be written as a layer-pseudospin spinor

$$|\psi\rangle = \eta_1|1\rangle + \eta_2 e^{i\varphi}|2\rangle, \qquad |\eta_1|^2 + |\eta_2|^2 = 1. \tag{S1}$$

In the ideal limit of vanishing layer separation and negligible interlayer tunnelling, the interaction is invariant under SU(2) rotations in layer space. A finite layer separation distinguishes intra- and interlayer Coulomb interactions and introduces an easy-plane anisotropy, while, in the absence of tunnelling, a relative U(1) symmetry remains. Interlayer exchange can spontaneously establish a finite off-diagonal coherence

$$\Delta_{12} = \langle c_1^\dagger c_2 \rangle \neq 0, \tag{S2}$$

thereby selecting a relative phase $\varphi$ and producing the familiar two-component QH exciton condensate [6-10].

For three active layers, the corresponding state is described by a three-component layer spinor,

$$|\psi\rangle = \eta_{\mathrm{t}}|\mathrm{t}\rangle + \eta_{\mathrm{m}} e^{i\varphi_{\mathrm{m}}}|\mathrm{m}\rangle + \eta_{\mathrm{b}} e^{i\varphi_{\mathrm{b}}}|\mathrm{b}\rangle, \tag{S3}$$

with

$$|\eta_t|^2 + |\eta_m|^2 + |\eta_b|^2 = 1 \tag{S4}$$

The overall phase is physically irrelevant, leaving two independent relative phases ($\varphi_{\mathrm{m}}$ and $\varphi_{\mathrm{b}}$). In the ideal zero-separation limit, this three-component manifold approaches an SU(3) layer symmetry, whereas the finite and unequal layer separations in the experimental devices explicitly introduce capacitive and exchange anisotropies. When interlayer tunnelling is negligible, however, two independent relative U(1) symmetries remain in the phase sector [20,21]. A fully three-component coherent state has finite coherence between every pair of active layers,

$$\Delta_{\mathrm{tm}} = \langle c_{\mathrm{t}}^\dagger c_{\mathrm{m}} \rangle, \qquad \Delta_{\mathrm{mb}} = \langle c_{\mathrm{m}}^\dagger c_{\mathrm{b}} \rangle, \qquad \Delta_{\mathrm{tb}} = \langle c_{\mathrm{t}}^\dagger c_{\mathrm{b}} \rangle, \tag{S5}$$

but the three coherence phases are not independent and satisfy the phase constraint

$$\arg(\Delta_{\mathrm{tm}}) + \arg(\Delta_{\mathrm{mb}}) - \arg(\Delta_{\mathrm{tb}}) = 0 \ (\mathrm{mod}\ 2\pi)\,. \tag{S6}$$

Such a state spontaneously breaks both relative U(1) symmetries and therefore supports two neutral collective phase degrees of freedom in the ideal coherent limit [20-22].

Pairwise coherent states arise as limiting cases in which only two layers participate in the active coherent sector while the third layer remains in an integer-filled incompressible state. For example, ($\eta_{\mathrm{b}} = 0$) reduces the three-component spinor to a top–middle EC, with analogous limits for top–bottom and middle–bottom coherence. We denote these configurations schematically as $|tm0\rangle$, $|t0b\rangle$ and $|0mb\rangle$, respectively, where 0 indicates that the corresponding layer does not participate in the active coherent subspace and does not necessarily imply zero total filling of that layer. The fully three-component configuration, in which all three amplitudes are finite, is denoted $|tmb\rangle$. Importantly, $|tmb\rangle$ represents a single three-component coherent state with two independent relative phases, rather than three independent pairwise condensates [20,21].

**Layer-resolved optical sensitivity**

The layer-resolved sensitivity of our optical measurements originates from the non-local, distance-dependent Coulomb coupling between the Rydberg excitonic states in monolayer $WSe_2$ and the electronic charge response of the nearby graphene layers [42,46,48,54]. Generally, the screening-induced renormalization of the $WSe_2$ excitonic resonance is governed by the momentum- and frequency-dependent charge susceptibility $\chi(q,\omega)$ of the proximate electronic system. The corresponding change in the screened Coulomb interaction can be written schematically as [48]

$$\Delta W(q,\omega) = [V(q)]^2 \chi(q,\omega), \qquad (S7)$$

where $V(q)$ is the interlayer Coulomb interaction between the $WSe_2$ sensing layer and the graphene layer. For two-dimensional layers separated by a distance $d$,

$$V(q) \propto \frac{e^{-\kappa q d}}{q}, \qquad (S8)$$

where $\kappa = \sqrt{\epsilon_{\parallel}/\epsilon_{zz}}$ accounts for the dielectric anisotropy of the surrounding medium. In the QH regime, LL occupation strongly modifies the charge response $\chi(q,\omega)$, including finite-momentum and low-energy contributions associated with LL excitations. The optical response therefore reflects a Coulomb-weighted charge response rather than solely the static compressibility of graphene. More importantly, the exponential factor continuously suppresses the coupling to more distant graphene sheets, with the degree of suppression depending on the momentum ($q$) sampled by a particular excitonic state. Graphene layers closer to $WSe_2$ therefore generally induce stronger modifications of the excitonic responses than more distant layers.

At the finite magnetic field used in the present measurements, the excited states are more appropriately regarded as magneto-Rydberg excitons. Their optical transition energies are determined jointly by the environmentally renormalized quasiparticle band gap, the screened electron–hole interaction and magnetic quantization [54-56]. Schematically,

$$E_{n,\tau}^{\mathrm{opt}}(B) = W + \varepsilon_n(B,W) + \Delta_{Z,\tau}, \qquad (S9)$$

where $E_g^{\mathrm{QP}}$ is the quasiparticle band gap, $W$ denotes the screened Coulomb interaction, $\varepsilon_n$ is the relative electron–hole eigenenergy including both Coulomb binding and magnetic confinement, and $\Delta_{Z,\tau}$ denotes the valley-dependent Zeeman contribution. Environmental screening modifies both $E_g^{\mathrm{QP}}$ and the exciton binding energy, which can partially compensate in the resulting optical transition energy [42,54]. For low-lying Rydberg states, magnetic confinement is a relatively small correction, whereas it becomes increasingly important at high principal quantum number ($N$) because the zero-field exciton radius increases rapidly while the binding energy decreases. The magnetic field therefore compresses high-lying Rydberg states and drives them progressively towards the magneto-exciton regime, where the cyclotron energy becomes an important energy scale [56]. We retain the *N*s notation to label these states according to their adiabatic connection to the zero-field Rydberg series. Even after magnetic compression, however, the higher-lying states remain substantially more spatially extended than the low-lying states at the magnetic fields investigated here. With increasing Rydberg order, the larger spatial extent of the exciton shifts its form factor towards smaller momenta, reducing the relative difference in Coulomb coupling to graphene layers separated by only a few nanometres. This accounts for the progressively weaker layer selectivity observed for higher Rydberg states.

The electronic state of graphene provides a second source of contrast. In the QH regime, LL occupation strongly modifies $\chi(q,\omega)$. When the Fermi level lies in an incompressible QH gap, the low-energy charge response is strongly suppressed over the momentum range relevant to the Rydberg exciton, reducing carrier-induced screening and spectral reconstruction and allowing well-defined neutral Rydberg resonances to recover. By contrast, when a LL is partially filled, the enhanced low-energy electronic response strongly modifies the excitonic spectrum. The optical signatures therefore reflect not only static compressibility but also finite-momentum and finite-frequency charge excitations of the Landau-quantized graphene system [44,45,48].

Coupling to partially filled graphene additionally produces interlayer exciton Fermi polarons. In this regime, the $WSe_2$ excitonic excitation is dressed by electronic excitations in the nearby graphene layer, giving rise to repulsive- and attractive-polaron branches on the higher- and lower-energy sides of the corresponding neutral-exciton resonance [45,57,58]. The strength of this spectral reconstruction depends on the Coulomb coupling to the participating graphene layer and therefore on its distance from the $WSe_2$ sensor [45,46,48]. Nearby layers generally produce stronger spectral-weight redistribution and more pronounced polaronic features, while the resonance energies can also exhibit layer-dependent shifts. We do not, however, interpret the exciton–polaron energy separation as a direct measure of the sensor–graphene distance or of a single microscopic interaction energy.

The neutral Rydberg and exciton-polaron responses therefore provide complementary forms of layer contrast. By constructing fixed-energy $\Delta R/R_0$ maps at several energies across the Rydberg-exciton and polaron resonances, we preferentially enhance different graphene-layer contributions and reconstruct the layer-resolved LL phase diagrams. The layer resolution used throughout this work thus results from the combined effects of the distance-dependent Coulomb form factor, the momentum- and frequency-dependent graphene charge response, the state-dependent magneto-Rydberg exciton form factor and the many-body exciton-polaron response.

**Construction of the ZLL phase diagram**

The optical measurements are based on dielectric sensing using Rydberg excitons in an adjacent monolayer $WSe_2$. As discussed above, these excited excitonic states are sensitive to the electronic charge response of the nearby graphene layers through long-range Coulomb coupling [41-48]. For each monolayer graphene sheet, the ZLL spans the filling range $-2 \leq \nu_i \leq 2$, with the fourfold spin–valley degeneracy further lifted into four symmetry-broken sublevels at the magnetic fields studied here. In the multilayer devices, the experimentally relevant ZLL manifold consists of the different combinations of these layer-resolved fillings. Within this manifold, the Rydberg-exciton and exciton-polaron features remain within a relatively narrow spectral window compared with the systematic evolution that occurs when higher orbital LLs are populated. The latter involves changes in the finite-momentum electronic screening and in the magneto-Rydberg exciton response and therefore generally produces a pronounced evolution of the excitonic resonance energy. We consequently restrict the phase diagrams discussed in this work to the filling range in which all participating graphene layers remain within their ZLL manifolds.

To visualize the ZLL structure, we construct fixed-photon-energy maps of the reflection contrast $\Delta R/R_0$ as a function of total filling factor $\nu_{\mathrm{tot}}$ and displacement field *D* (See more in Extended Data Fig. 2 and 4). The photon energies are chosen to intersect selected Rydberg-exciton or exciton-polaron resonances that exhibit strong sensitivity to the ZLL occupation. Depending on the device and the desired layer contrast, we use several different probe energies, including the 2s and 3s exciton resonances and lower-energy exciton-polaron features. Because these resonances possess different spatial and many-body sensitivities, maps obtained at different photon energies preferentially emphasize different graphene layers or different components of the electronic charge response. The resulting fixed-energy maps should therefore be regarded as optical representations of the ZLL charge-response and incompressibility structure rather than direct thermodynamic compressibility measurements.

For the dual-gated devices, the total carrier density is determined from the top- and bottom-gate voltages according to

$$n_{\mathrm{tot}} = \frac{C_{\mathrm{tg}}V_{\mathrm{tg}} + C_{\mathrm{bg}}V_{\mathrm{bg}}}{e} - n_0, \qquad (S10)$$

and the average perpendicular displacement field is

$$D = \frac{C_{\mathrm{tg}}V_{\mathrm{tg}} - C_{\mathrm{bg}}V_{\mathrm{bg}}}{2\epsilon_0} - D_0, \qquad (S11)$$

where $C_{\mathrm{tg}}$ and $C_{\mathrm{bg}}$ are the top- and bottom-gate capacitances per unit area, respectively, *e* is the elementary charge, and $n_0$ and $D_0$ account for residual charge doping and electrostatic offset. Here *D* is expressed in electric-field units mV/nm.

At finite magnetic field, the total filling factor is obtained from $\nu_{\mathrm{tot}} = \frac{n_{\mathrm{tot}}h}{eB}$, with the density calibration cross-checked against the experimentally resolved integer QH states. In the double- and triple-layer devices, the gate voltages determine the total carrier density, whereas the distribution of charge among the individual graphene layers depends additionally on *D*, the applied interlayer bias $V_{\mathrm{bias}}$ and the electrostatic interactions between the layers. The individual layer fillings $\nu_{\mathrm{t}}$, $\nu_{\mathrm{m}}$ and $\nu_{\mathrm{b}}$ used in the phase diagrams are assigned from the layer-resolved optical trajectories together with the electrostatic model described below.

**Simulation of the ZLL phase diagram**

To simulate the zero-Landau-level (ZLL) crossing phase diagram of the dual-gated multilayer graphene device, we describe the carrier distribution by combining electrostatic charge balance with a phenomenological chemical-potential function $\mu_i(\nu)$ for each graphene layer, where $i = t, m, b$ labels the graphene layer within the device under consideration.

In our minimal model, each chemical-potential function (see Extended Data Fig. 6) includes large cyclotron gaps at the fillings ($\nu = \pm 2, \pm 6, \pm 10, \ldots$) where the LL orbital index changes [59,62], smaller quantum Hall ferromagnetic (QHFM) gaps ($\nu = -1, 0, 1$) [33-40,59], and interaction-driven negative compressibility inside the ZLL [59-61]. The dominant single-particle energy scale is set by the graphene LL spacing [62]

$$E_1 = v_F \sqrt{2e\hbar B}, \qquad (S12)$$

where $v_F$ is the graphene Fermi velocity. While the cyclotron gaps are controlled by a common LL energy scale, the QHFM gaps are allowed to vary slightly among the layers to phenomenologically account for differences in dielectric environment and residual disorder (see Extended Data Fig. 6 for more details). Once determined, the layer-dependent chemical-potential parameters are kept fixed throughout the calculations presented here.

With the phenomenological chemical-potential function defined above, we first consider the dual-gated DLG device. The two graphene layers are electrically grounded and therefore act as carrier reservoirs. The equilibrium condition is that the electrochemical potential of each graphene layer is pinned to the common reservoir value. Denoting the electrical potentials of the top and bottom graphene layers by $\phi_t$ and $\phi_b$, and their chemical potentials by $\mu_t$ and $\mu_b$, we write

$$\tilde{\mu}_t = \mu_t - e\phi_t = 0, \tilde{\mu}_b = \mu_b - e\phi_b = 0. \qquad (S13)$$

Equivalently,

$$\phi_t = \mu_t / e, \phi_b = \mu_b / e. \qquad (S14)$$

Let $c_t$, $c_i$, and $c_b$ be the geometric capacitances per unit area between the top gate and top graphene, between the two graphene layers, and between the bottom graphene and bottom gate, respectively. For a dielectric stack, the capacitance per unit area is given by the series combination

$$c^{-1} = \sum_j \frac{d_j}{\epsilon_0 \epsilon_j}, \qquad (S15)$$

where $d_j$ and $\epsilon_j$ are the thickness and relative dielectric constant of the $j$-th dielectric layer. The carrier densities of the top and bottom graphene layers are written in terms of their filling factors as

$$n_t = \nu_t \frac{eB}{h}, n_b = \nu_b \frac{eB}{h}, \qquad (S16)$$

with $B$ the magnetic field. The electrostatic charge-balance equations then read

$$en_t = c_t(V_{tg} - \phi_t) + c_i(\phi_b - \phi_t), \qquad (S17)$$

$$en_b = c_b(V_{bg} - \phi_b) + c_i(\phi_t - \phi_b). \qquad (S18)$$

Substituting Eq. (S14) into Eqs. (S17) and (S18), one obtains two coupled self-consistency equations for $\nu_t$ and $\nu_b$,

$$en_t = c_t(V_{tg} - \mu_t/e) + c_i(\mu_b/e - \mu_t/e), \qquad (S19)$$

$$en_b = c_b(V_{bg} - \mu_b/e) + c_i(\mu_t/e - \mu_b/e), \qquad (S20)$$

where $\mu_t = \mu_t(\nu_t)$ and $\mu_b = \mu_b(\nu_b)$.

The same framework can be extended straightforwardly to a dual-gated TLG structure. For notational simplicity, the labels $t$ and $b$ are used for the top and bottom layers in both device geometries. However, the corresponding filling factors and chemical potentials are determined independently in the DLG and TLG calculations. In this case, one introduces three-layer filling factors $\nu_t, \nu_m, \nu_b$, three-layer chemical potentials $\mu_t, \mu_m, \mu_b$, and three-layer electrical potentials $\phi_t, \phi_m, \phi_b$. In the TLG measurements, top layer graphene is connected to a biased voltage of $V_{\text{bias}}$, while the middle and bottom graphene layers are still treated as carrier reservoirs connected to the common ground, so that

$$\tilde{\mu}_t = \mu_t - e\phi_t = -eV_{bias}, \qquad (S21)$$

$$\tilde{\mu}_m = \mu_m - e\phi_m = 0, \qquad (S22)$$

$$\tilde{\mu}_b = \mu_b - e\phi_b = 0. \qquad (S23)$$

The carrier density in each layer is

$$n_i = \nu_i \frac{eB}{h}, i = t, b, m. \qquad (S24)$$

The electrostatic model is then determined by the full capacitance network connecting the three graphene layers to each other and to the two external gates. Writing the geometric capacitances between adjacent conducting planes as $c_{ij}$, the charge on each graphene layer is obtained from the sum of all capacitively induced charges from its neighboring electrodes and graphene sheets. This produces three coupled self-consistency equations of the form

$$en_i = \sum_j c_{ij}\left(\phi_j - \phi_i\right) + c_{ig}\left(V_g - \phi_i\right) \qquad (S25)$$

where the appropriate gate term is included only for the outermost graphene layers. After substituting $\phi_i$ in terms of the $\mu_i(\nu_i)$ and the applied bias through Eqs. (S21)–(S23), one obtains a set of three coupled nonlinear equations for $\nu_t, \nu_m, \nu_b$.

In this way, the triple-layer problem is a natural extension of the double-layer model: the electrostatic charge-balance equations are enlarged from two coupled equations to three, while the essential physical ingredients remain the same, namely LL jumps, ZLL symmetry-breaking gaps, and negative compressibility in the ZLL. These ingredients determine how the applied gate and bias voltages redistribute carriers among the different graphene layers and thereby generate the ZLL crossing phase diagram.

## Figures

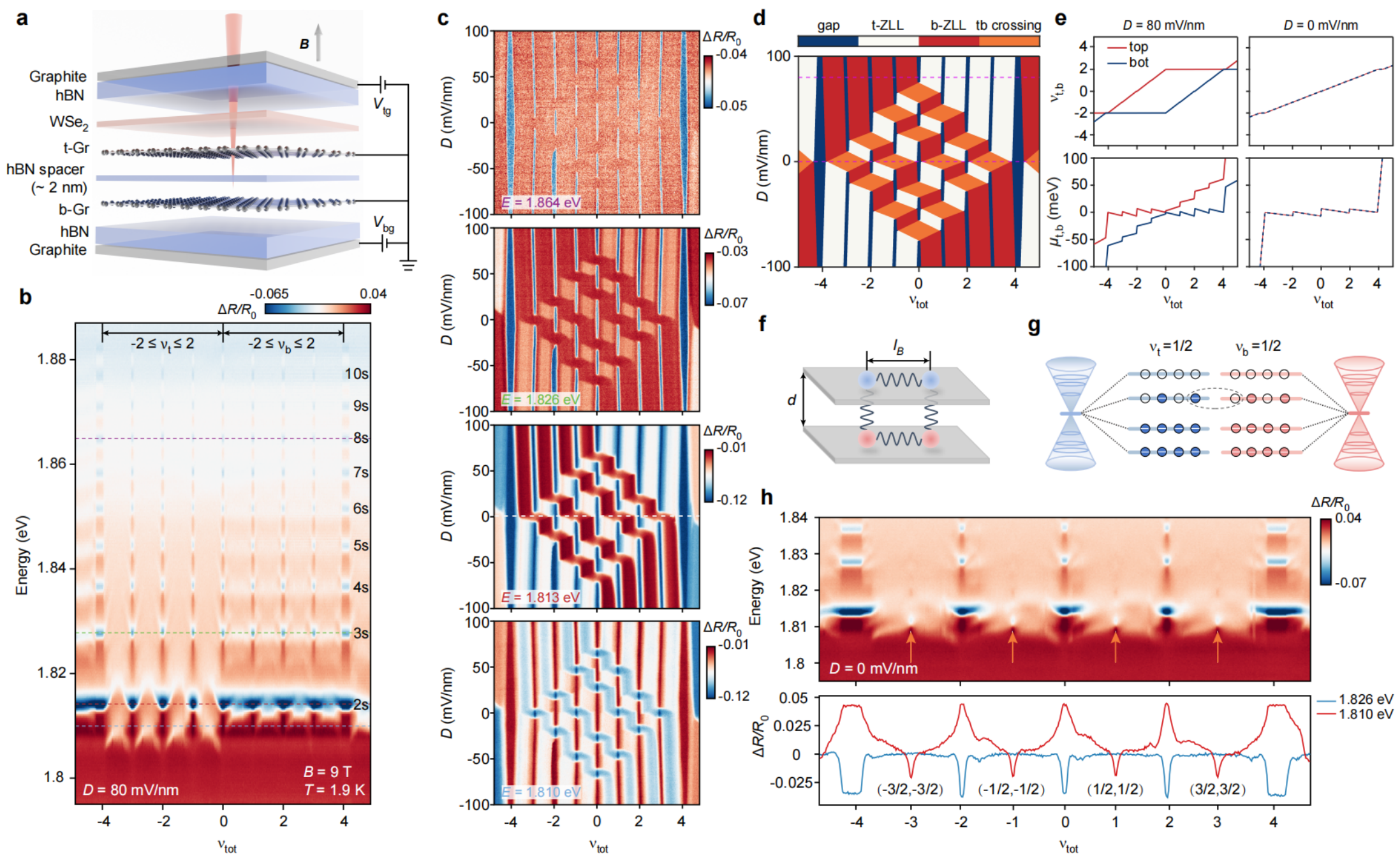

**Figure 1. Layer-resolved Rydberg exciton sensing spectroscopy of interlayer two-component QH coherence in double-layer graphene (DLG). a,** Schematic of the DLG/$WSe_2$ heterostructure, with the t-Gr and b-Gr (two graphene monolayers) separated by an approximately 2 nm hBN spacer. A monolayer $WSe_2$ is placed above t-Gr and serves as an optical sensor. The total filling factor $\nu_{\mathrm{tot}}$ and displacement field $D$ are controlled by the top- and bottom-graphite gates ($V_{\mathrm{tg}}$ and $V_{\mathrm{bg}}$, respectively) under an out-of-plane magnetic field $B$. **b,** Reflection-contrast spectrum $\Delta R/R_0$ as a function of photon energy ($E$) and $\nu_{\mathrm{tot}}$, measured at 80 mV/nm, showing the sequential filling of the ZLLs of the two graphene layers ($\nu_{\mathrm{t}}$ and $\nu_{\mathrm{b}}$, respectively). The $WSe_2$ Rydberg exciton resonances from 2s to 10s are indicated. **c,** Phase diagrams extracted from the reflection contrast at fixed photon energies of 1.864, 1.826, 1.813 and 1.810 eV (highlighted by the dashed lines in **b**), corresponding respectively to the $WSe_2$ 8s, 3s and 2s exciton resonances and the 2s attractive-polaron branch. **d,e,** Simulated ZLL crossing phase diagram (**d**) and representative linecuts at $D$ = 80 and 0 mV/nm (**e**) of DLG. Red and white regions denote the symmetry-broken ZLLs of t-Gr and b-Gr, respectively, orange regions mark crossings between top- and bottom-layer ZLL branches, and blue regions denote the QH gaps. **f,** Real-space schematic of an interlayer coherent QH state with two characteristic length scales: interlayer separation $d$ and magnetic length $l_B$. **g,** Schematic of interlayer electron-hole pairing at $(\nu_{\mathrm{t}}, \nu_{\mathrm{b}})$ = (1/2, 1/2). The active half-filled ZLL sublevels of the t-Gr and b-Gr are brought into resonance and can form interlayer EC. **h,** Upper panel, $\Delta R/R_0$ spectrum measured versus $\nu_{\mathrm{tot}}$ along $D$ = 0. Lower panel, linecuts extracted at 1.826 eV (3s exciton energy) and 1.810 eV. The latter energy is selected to track the red-shifted spectral response associated with the additional incompressible states at odd values of $\nu_{\mathrm{tot}}$ (highlighted by the orange arrows), which are corresponding respectively to the layer fillings $(\nu_{\mathrm{t}}, \nu_{\mathrm{b}}) = (\pm 3/2, \pm 3/2)$ and $(\pm 1/2, \pm 1/2)$. These features occur between the larger

neighboring regular QH gaps and are consistent with EC gap opening at top–bottom ZLL crossings. Unless otherwise specified, all measurements were performed at $B = 9$ T and $T = 1.9$ K.

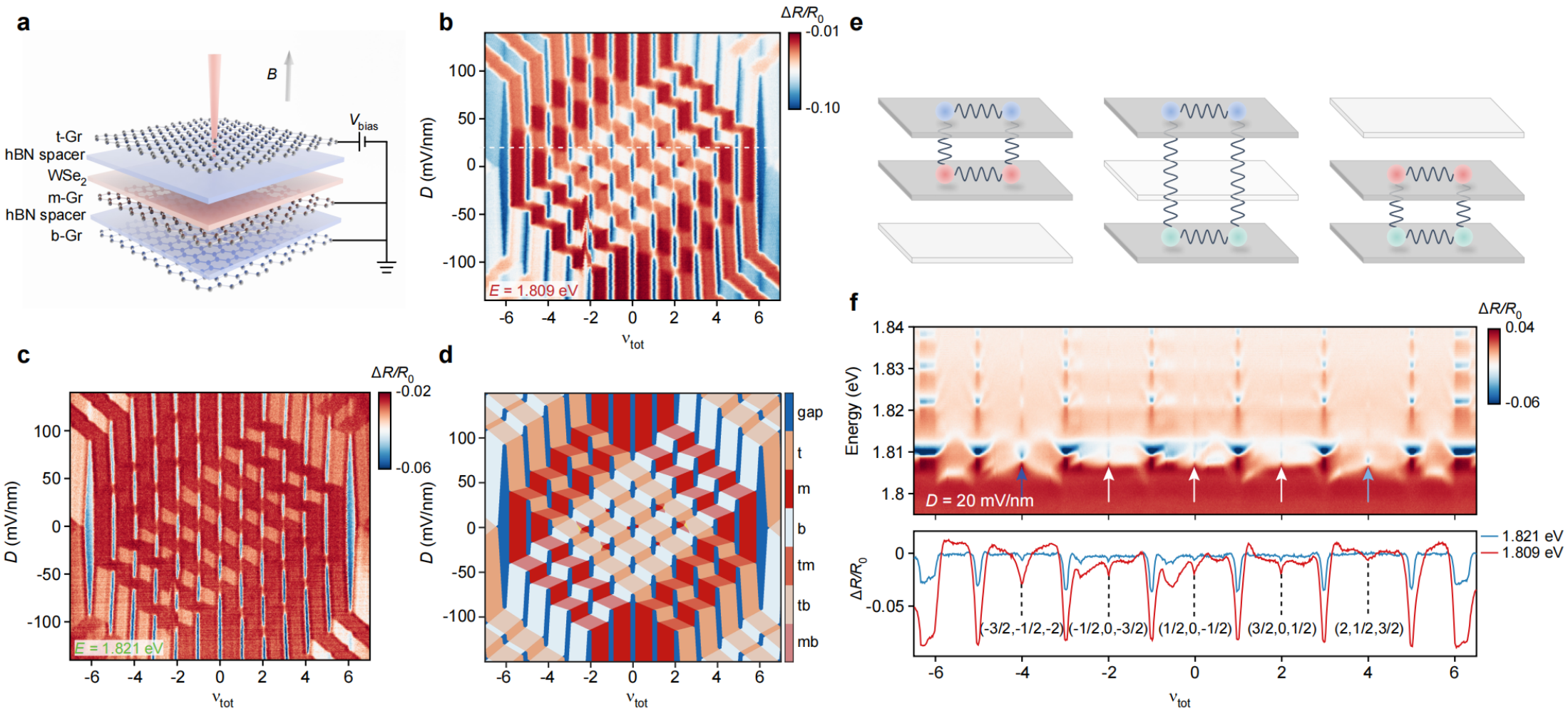

**Figure 2. Pairwise interlayer coherent states in triple-layer graphene (TLG). a,** Schematic of the TLG/$WSe_2$ heterostructure (not showing the outside gating graphite and hBN). Three monolayer graphene sheets, denoted t-Gr, m-Gr and b-Gr, are separated by two hBN spacers with thicknesses 1.3 nm and 2 nm, respectively. A bias voltage $V_{\text{bias}}$ can be applied to the t-Gr, while m-Gr and b-Gr are grounded. **b,c,** Layer-resolved ZLL phase diagrams constructed from the reflection contrast at fixed photon energies of 1.809 eV (**b**) and 1.821 eV (**c**), corresponding to the 2s and the 3s exciton resonance, respectively. The horizontal dashed line in **b** marks $D$ = 20 mV/nm, along which the spectrum and linecuts in **f** are measured. **d,** Simulated ZLL crossing map of TLG. The colors encode the individual layer-resolved ZLL regions and the regions in which sub-ZLL branches from two layers cross. The calculation describes the electrostatic evolution and pairwise crossing structure of the layer-resolved LLs. **e,** Schematics of the three possible pairwise interlayer-coherent states in TLG: top–middle, top–bottom and middle–bottom coherence, from left to right. **f,** Upper panel, reflection-contrast spectrum measured along $D$ = 20 mV/nm. Lower panel, linecuts extracted at of 1.821 eV and at 1.809 eV. Vertical dashed lines indicate the corresponding layer-resolved filling configurations, written in the order $(\nu_t, \nu_m, \nu_b)$. The additional incompressible features occur when two layers are at half-integer filling while the third remains at integer filling, consistent with pairwise interlayer EC states involving all three possible layer combinations. These states include $(-3/2, -1/2, /2)$ and $(2, 1/2, 3/2)$, involving top–middle (indicated by the dark-blue arrow) and middle–bottom crossings (light-blue arrow), respectively, whereas $(-1/2, 0, -3/2)$, $(1/2, 0, 1/2)$ and $(1/2, 0, 3/2)$ involving top–bottom crossings (white arrows).

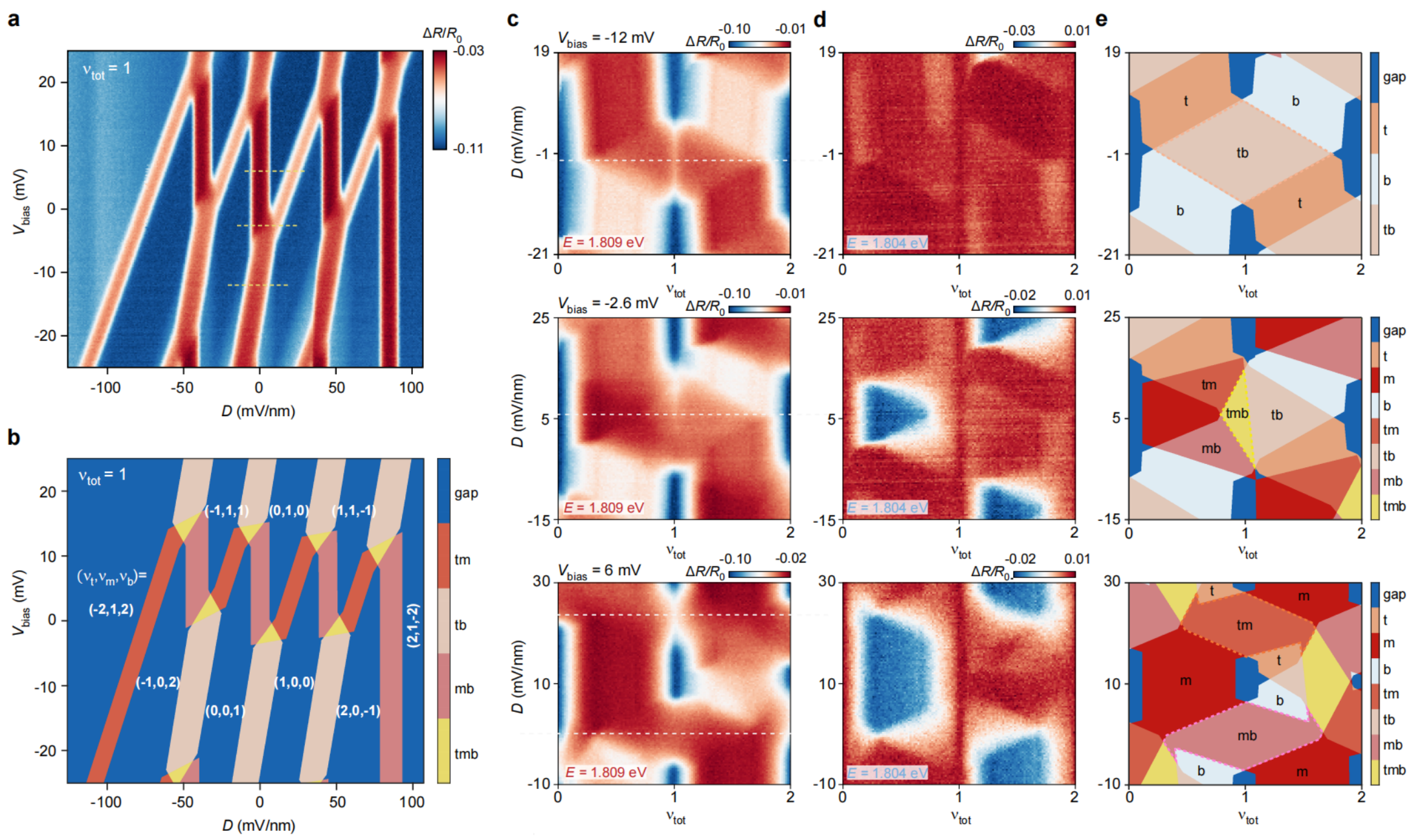

**Figure 3. Bias-tuned evolution from pairwise to three-layer LL crossings in TLG.** **a,** $V_{\mathrm{bias}}$ and $D$ dependent phase diagram of the TLG at fixed total filling $\nu_{\mathrm{tot}} = 1$, extracted from the reflection contrast at the fixed 2s exciton energy of 1.809 eV. The bias voltage $V_{\mathrm{bias}}$ is applied to t-Gr. Blue regions correspond to regular QH gaps, whereas the red-toned branches between adjacent blue domains mark interlayer LL crossings. From the darkest to the lightest red, these regions are assigned to the middle–bottom (mb), top–bottom (tb) and top–middle (tm) coherent channels, respectively. Multiple sets of such interlayer crossings evolve continuously as $V_{\mathrm{bias}}$ and $D$ are varied. The three yellow dashed lines indicate representative set of bias-tunable crossings selected for detailed investigation in **c–e**, illustrating their evolution from pairwise crossings to a three-layer LL crossing regime. **b,** Simulation of the $V_{\mathrm{bias}}$- and $D$-dependent evolution at $\nu_{\mathrm{tot}} = 1$. The corresponding layer filling configurations in the blue regions are explicitly labeled. The simulation reproduces the experimentally observed evolution from pairwise crossings to three-layer crossing regimes at the junctions (yellow triangles, labeled as tmb). **c,d,** Zoomed-in ZLL phase diagrams at $V_{\mathrm{bias}} = -12$, $-2.6$, and $6$ mV, constructed from the reflection contrast at fixed energies of 1.809 eV (**c**) and 1.804 eV (**d**). The 1.809 eV maps show the overall layer-resolved ZLL structure, whereas the 1.804 eV maps provide complementary layer selectivity and enhance the visibility of the middle-layer LLs, represented by the pronounced blue features in **d**. The white dashed lines in **c** indicate the displacement field cuts used to obtain the spectra in Fig. 4a–d. **e,** Simulated layer-resolved ZLL crossing maps at the same three bias voltages as in **c**,**d**. The dashed boundaries highlight several representative crossing regions labeled with the layer-participation configuration.

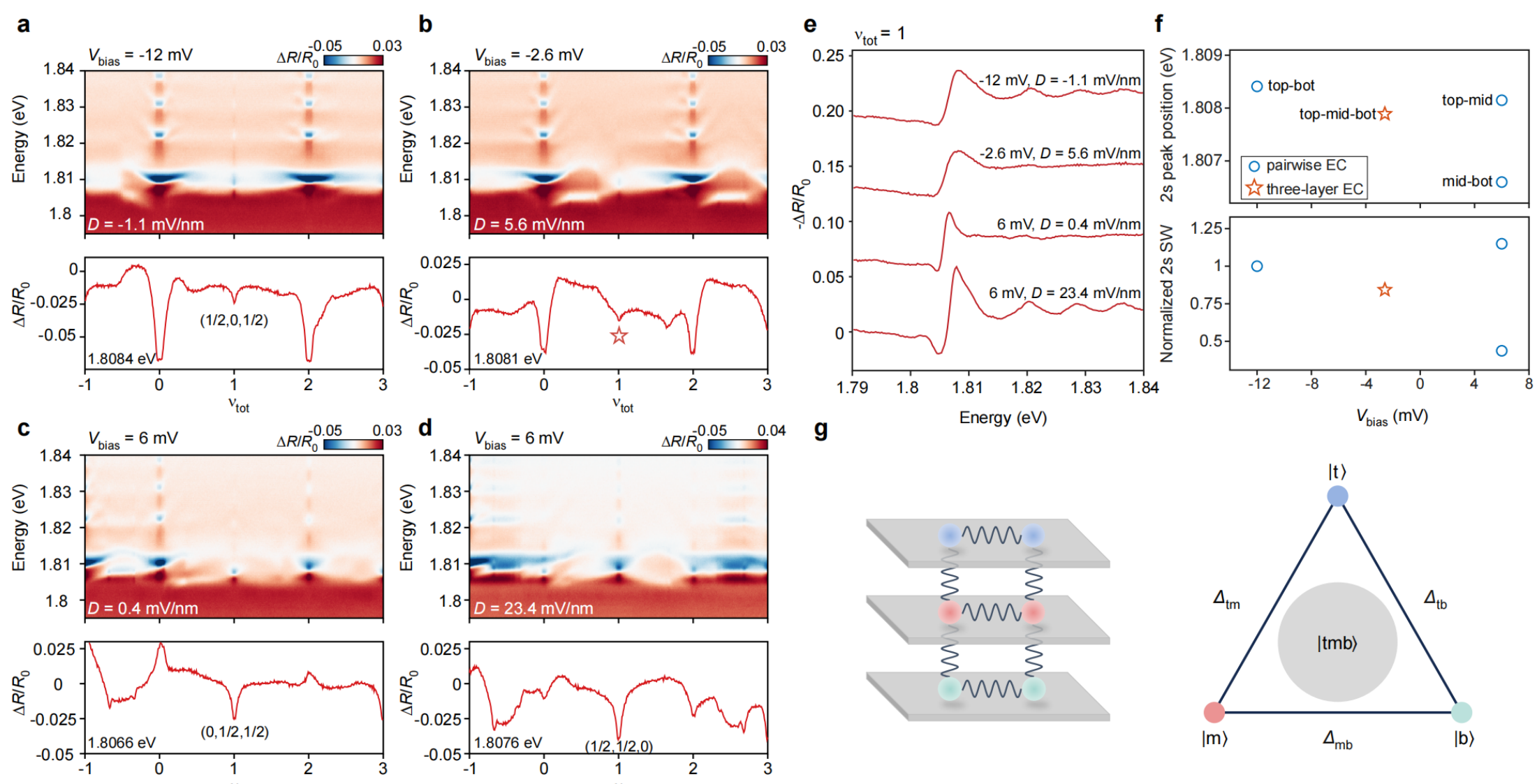

**Figure 4. Signatures of three-component QH coherence. a–d,** Reflection-contrast spectra measured along the four displacement field cuts indicated by the white dashed lines in Fig. 3c. The upper panels show the spectra as functions of photon energy $E$ and $\nu_{tot}$, and the lower panels show linecuts at the indicated photon energies, selected to maximize the contrast of the corresponding incompressible features at $\nu_{tot} = 1$. Panels **a–d** correspond, respectively, to top–bottom pairwise coherence at $V_{bias} = -12$ mV, the three-layer crossing regime at $V_{bias} = -2.6$ mV, and the middle–bottom and top–middle pairwise coherent states at $V_{bias} = 6$ mV. The orange star in **b** marks the unique incompressible feature at $\nu_{tot} = 1$, where the relevant ZLLs from all three layers approach simultaneous degeneracy. **e,** Reflection contrast spectra at $\nu_{tot} = 1$, extracted from the four states shown in **a–d** and vertically offset for clarity. The spectra compare the $WSe_2$ excitonic response of the top–bottom, three-layer, middle–bottom and top–middle configurations. **f,** Extracted 2s peak position (upper panel) and normalized 2s spectral weight (SW, lower panel) for the states in **a–d**, plotted as functions of $V_{bias}$. Blue circles denote the three pairwise coherent states, whereas the orange star marks the incompressible feature at the three-layer crossing. **g,** Schematic illustrating the construction of three-component interlayer coherent EC $|tmb\rangle$. Left, real-space picture of coherent coupling among all three graphene layers. Right, pseudospin or order-parameter representation in the layer basis $|t\rangle$, $|m\rangle$ and $|b\rangle$. The three pairwise phase coherences are labeled by $\Delta_{tm}$, $\Delta_{tb}$, and $\Delta_{mb}$, respectively.